# Giant resonance and anomalous quality factor scaling in coupled resonator optical waveguides at the degenerate band edge

Mohamed Y. Nada, Mohamed A. K. Othman, Ozdal Boyraz, and Filippo Capolino

*Department of Electrical Engineering and Computer Science, University of California, Irvine, CA 92697, USA*

We propose a novel scheme for enhancing the quality factor of coupled resonators optical waveguides (CROWs) when operating near a degenerate band edge (DBE). A DBE is a four-mode exceptional point of degeneracy (EPD) occurring when four Bloch eigenmodes coalesce providing a resonance condition with a giant enhancement in fields. We report an unprecedented scaling law of quality factor of CROWs when operating at the DBE, even in the presence of losses and structural perturbations. Remarkably, the $Q$ factor of the proposed CROW can be engineered to exceed that of a single ring resonator having a diameter equal to the CROW length, hence having an overall strong area reduction. The findings reported in this letter are critical for enhancing field's amplitudes to giant levels and the $Q$ factor of ring resonators and are very beneficial for various applications including four wave mixing, $Q$ switching, lasers, and highly sensitive sensors.

## 1. INTRODUCTION

High quality ($Q$) factor microcavities provide a practical testbed for new advances in fundamental sciences, in particular biological and chemical sensing. Implementation of such high $Q$ factor microcavities has been a classical contest in the optics realm [1]. Thanks to state of the art nanofabrication techniques such high $Q$ cavities have been ubiquitous for various on-chip photonic devices. A recent burgeoning aspect in high $Q$ cavity design is the concept of slow light in which a field in an optical guiding system possesses a group velocity much lower than the velocity of light in vacuum $c$ [2,3]. The proliferation of slow light has spawned many intriguing aspects in light manipulation and transport for which nonlinearities (higher harmonic generation, wave mixing, etc.) [4], and gain/absorption [5] among other features can be significantly enhanced. In this letter we demonstrate a fundamentally novel approach for realizing high $Q$ factor microcavities using a special kind of engineered slow-light through eigenmode dispersion and degeneracy conditions [6]. Particularly, slow light resonance occurring in the vicinity of the band edge of periodic structures is intimately linked to degeneracies of Bloch eigenmodes. This degeneracy condition occurs when wave propagating eigenvectors coalesce. A degenerate band edge (DBE) [7–10] arises when four Bloch eigenstates (eigenvalues *and* eigenvectors) coalesce into a single one in periodic structures supporting multiple polarization eigenstates that are periodically mixed. This has led to many interesting physical processes in optics [10,11], and microwaves [12,13]. The concept of exceptional point [14,15] has already received a surge of interest in recent years. Parity-Time symmetry and the DBE are two distinct classes of systems with exceptional points, where in the latter class the guiding structure has neither losses nor gain. The DBE investigated here is a fourth order exceptional point of degeneracy (EPD). Although degeneracy condition is an exact mathematical condition that can only be achieved when one parameter is rigorously met, we show here that the desired performance related to these degeneracies can still be detected even when structural perturbations occur, as also seen in a recent microwave experiment [13]. Here we propose an optical platform based on coupled resonator optical waveguide (CROW) design introduced by Yariv et al. in [16]. The proposed design (see Fig. 1) leads to observing the DBE and large $Q$ factors higher than $10^5$ in relatively small structures even in the presence of realistic material loss and perturbations existing due to potential microfabrication process tolerances. This shows great promise for realizing high $Q$ factor compared to analogous designs investigated in [17,18]. Interestingly, we demonstrate that such CROW can possess higher $Q$ factors than that of a single microring resonator having a diameter equals the length of the CROW and same material loss.

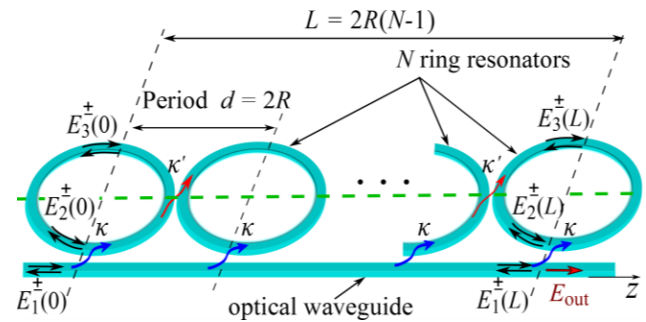

Fig. 1. The loaded CROW consists of a chain of $N$ ring resonators of radius $R$, coupled to each other with field coupling coefficient $\kappa'$. They are side-coupled to a rectangular waveguide with $\kappa$. The structure is periodic in the $z$-direction with a period $d = 2R$. The figure also shows the field amplitudes defined at points $z = 0$ and $z = L$.

## 2. ANOMALOUS SCALING OF THE $Q$ FACTOR AT THE DBE

Let us consider the CROW in Fig. 1 designed to exhibit the DBE. A DBE is a condition upon which four eigenvectors representing wave propagating in the periodic structure coalesce. This could only occur in structures supporting multiple polarizations guided and coupled through the periodic waveguide (such as the anisotropic/birefringent layers [5]), or in coupled waveguides such as the CROW in Fig. 1. The proposed CROW is composed of a chain of ring resonators coupled to each other via coupling coefficient $\kappa'$, and side-coupled to a uniform waveguide with another coupling coefficient $\kappa$. The waveguide and the rings have effective refractive indices $n_r$ and $n_w$, respectively, and we assume single transverse mode propagating in each waveguide. Each ring resonator radius is $R$ hence the periodic CROW has period $d = 2R$. Note that we ignore the gap dimensions between rings as well as the ring thickness for simplicity as was done in [2]. To explore the unique modal characteristics of this CROW, we proceed by representing wave propagation along z using complex field amplitudes that are defined as shown in Fig. 1. As such, there exist at any point z three complex field amplitudes that propagate in the positive z-direction, namely $E_1^+(z)$, $E_2^+(z)$, and $E_3^+(z)$, and they are described by a three-dimensional vector $\mathbf{E}^+(z) = \begin{bmatrix} E_1^+(z) & E_2^+(z) & E_3^+(z) \end{bmatrix}^T$. Analogously, three field amplitudes at the same point z represent wave that propagates along the negative z-direction with a field amplitude vector $\mathbf{E}^-(z) = \begin{bmatrix} E_1^-(z) & E_2^-(z) & E_3^-(z) \end{bmatrix}^T$. To quantitatively analyze the wave dynamics, we assume a state vector composed of the six field amplitude components $\mathbf{\psi}(z) = [(\mathbf{E}^+(z))^T \ (\mathbf{E}^-(z))^T]^T$. Then, we utilize coupled mode theory [2,19] and a 6×6 transfer matrix formalism to investigate the evolution of the state vector along the CROW and derive the eigenmode characteristics. Accordingly, at any frequency, there exist up to six Bloch eigenmodes guided by the CROW. Yet at some particular frequencies some eigenmodes coalesce in both their wavenumber and eigenvectors. We generalize the theory of CROW [2] to the case shown in Fig. 1 and obtain the dispersion relation of the eigenmodes of the system, namely $D(k,\omega)=0$, where $k$ is the Bloch wavenumber along $z$ and $\omega$ is the angular frequency. There exist six Bloch wavenumber solutions and they obey the symmetry such that $k$ and $-k$ are both solutions. It is important to point out that conventional CROWs made of only a chain of coupled ring resonators have a structural symmetry in which their modes exhibit only a regular band edge (RBE) as shown in [2]. Introducing coupling to the straight waveguide as in the geometry depicted in Fig. 1 breaks the structural symmetry and in turn facilitates the observation of a DBE, i.e., a fourth order degeneracy. Symmetry here is defined with respect to a plane cutting the rings in half and perpendicular to the plane containing the rings as shown in Fig. 1 with a horizontal dashed line. The DBE is found by proper tuning of the coupling parameters, effective refractive indices and radius of the rings. Although there are many possible points in the parameter space of the CROW that realize DBEs, we focus on some designs to demonstrate important resonance characteristics described in the following. The DBE wavelength is chosen close to $\lambda_d = 2\pi c / \omega_d = 1550$ nm in all the subsequent analysis. We consider three different designs of the unit cell of the periodic CROW; whose parameters are given in Table. 1. The reported coupling coefficients and effective refractive indices of the CROWs under consideration can be readily implemented using silicon optical ridge waveguides as in [20,21]. The dispersion diagram of the three designs of the DBE CROW is depicted in Fig. 2(a); near the DBE wavelength (only real branches are shown in the range $kd \in [0, 2\pi]$). Note that the dispersion relation in the vicinity of DBE frequency is approximated by $(1 - \omega / \omega_d) \cong \zeta (k / k_d - 1)^4$ where $\omega_d$ is the DBE angular frequency and $k_d = \pi / d$ is the wavenumber at the band edge.

The parameter $\zeta$ dictates the flatness of the dispersion relation, i.e., the value of the fourth derivative $d^4\omega/dk^4$ at the DBE. Smaller values of $\zeta$ indicate flatter dispersion at the DBE, and it plays a very important role in realizing higher $Q$ factors. For that aim, we investigate the transmission properties of the finite CROW with $N$ cascaded rings and examine the resonances near the DBE. In Fig. 2(b) we show the transmission coefficients of the three CROW designs made of $N = 16$ coupled rings. The transmission coefficient is defined as $T_F = |E_{\text{out}} / E_1^+(0)|$ where $E_{\text{out}}$ is the field amplitude exiting the waveguide from the right, while $E_1^+(0)$ is the field amplitude representing the excitation of the waveguide. The transmission peaks of the CROW have a narrow spectral width when the frequency approaches the DBE as seen from Fig. 2(b) (Note that the straight waveguide itself does not have discontinuities). The transfer function $T_F$ also has a unity magnitude for such resonance in a lossless CROW. Moreover, in Fig. 2(c) one can observe how the resonance angular frequencies closest to DBE denoted by $\omega_{r,d}$ evolve as the length of the CROW increases; for the three different designs having various $\zeta$s. The DBE resonance angular frequency $\omega_{r,d}$ is getting closer to $\omega_d$ either by increasing $N$ or decreasing $\zeta$. Such trend follows the asymptotic formula $\omega_{r,d} / \omega_d \approx 1 - \zeta / N^4$ as discussed in [7].

**TABLE 1: THREE DESIGNS OF DBE CROW UNIT CELL, AND THE CORRESPONDING VALUES OF THE $Q$ FACTOR OF A SINGLE RING RESONATOR IN BOTH LOSSY ($Q_{0,\text{loss}}$) AND LOSSLESS ($Q_0$) CASES**

| # | $\kappa$ | $\kappa'$ | $n_w$ | $n_r$ | $R$(μm) | $\zeta$ | $Q_0$ | $Q_{0,\text{loss}}$ |
|---|---|---|---|---|---|---|---|---|
| 1 | 0.7 | 0.09 | 2.51 | 2.48 | 50 | 0.001 | $4.7\times10^3$ | $4.5\times10^3$ |
| 2 | 0.25 | 0.1 | 2.4 | 2.51 | 10 | 0.06 | $1\times10^4$ | $9.3\times10^3$ |
| 3 | 0.25 | 0.1 | 2.42 | 2.5 | 50 | 0.01 | $5\times10^4$ | $3.5\times10^4$ |

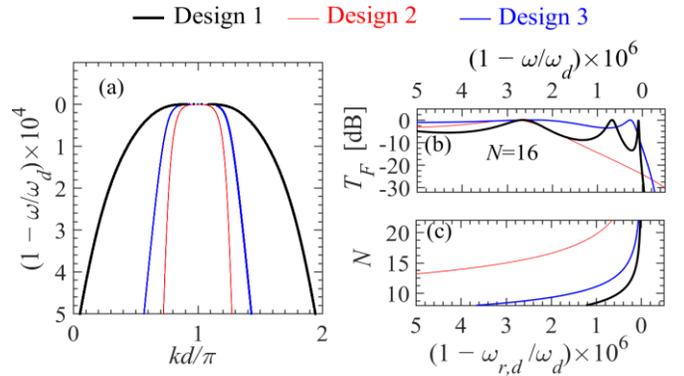

Fig. 2. (a) The Bloch wavenumber dispersion showing the propagating modes (purely real k) of the three designed CROWs with parameters given in Table. 1. (b) The transfer function $T_F$ calculated near the DBE resonance $\omega_{rd}$ for the three CROW lossless designs with $N$=16. (c) Trajectory of the DBE resonance frequency of the lossless CROWs for different $N$ that follows the trend $\omega_{r,d} / \omega_d \approx 1 - \zeta / N^4$. For the sake of clarity, note the normalization of the angular frequency axes in the three plots.

Now, we analyze the scaling of the $Q$ factor with the length of a CROW shown in Fig. 1. The loaded $Q$ factor of the cavity (connected to the straight waveguide at both ends) versus the number of rings $N$ is shown in Fig. 3(a) for the three CROW designs. From here onward, the loaded quality factor of the DBE resonator is referred to as "$Q$ factor" and is calculated through the group delay, obtained as discussed in [10]. The $Q$ factor is evaluated at the DBE resonance angular frequency $\omega_{r,d}$ for each respective design, see

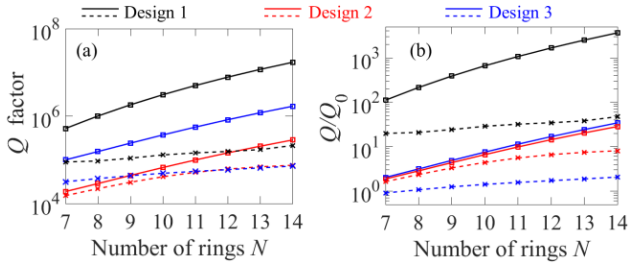

Fig. 3. (a) The loaded quality factor of three designed CROWs in the ideal lossless (markers on solid lines) and lossy waveguide (markers on dashed lines) cases. (b) The loaded $Q$-factor normalized to the $Q_0$ of a single loop, for each design. The values of $Q_0$ and $Q_{0,\text{loss}}$ are given in Table. 1. The solid curves represent the lossless $Q$ fitted by the equation $aN^5+b$, that is in perfect agreement with the simulated $Q$ represented with square markers.

Fig. 2(c). We observe the general trend for lossless DBE structures in which the $Q$ factor is fitted by $aN^5 + b$ (see also [7,10,17]) where the fitting parameters $a$ and $b$ are different for the three designed CROWs. It is important to point out that the growth as $Q \propto aN^5$ represents an unprecedented scaling of $Q$ factor for CROWS. The parameter $a$ is inversely proportional to the dispersion fitting parameter $\zeta$, in the sense that the product of $a$ and $\zeta$ is approximately constant, i.e., $a\zeta \simeq$ constant for the three cases under consideration and equals to ~0.03. The later observation is inherently related to the fact that the quality factor is inversely proportional to the group velocity of the wave $v_g$ in the constitutive periodic structure (i.e. $Qv_g$ = constant) as discussed in [3], and $v_g$ is in turn proportional to $\zeta$ in the vicinity of the DBE frequency. Therefore, realizing smaller values of $\zeta$ (meaning flatter dispersion) leads to higher quality factors. In essence, this would also entail an increase in the local density of states [9] when wavenumber dispersion is flatter. The anomalous scaling law for large $N$ applies to the lossless structures, whereas the effect of losses is described next.

We account for radiation and dissipative losses by incorporating the attenuation constant of the waveguide and ring resonators. We assume that dissipative losses for silicon are 3.7 dB/cm and radiation losses are 0.005 dB/turn due to bending as given in [22]. In Fig. 3(a) we show the $Q$ factor for lossy CROWs. Losses cause a saturation effect in the anomalous $Q$ factor scaling law. Indeed, the $Q$ factor grows by increasing the number of rings $N$ but after a certain length the growth ceases to have the $N^5$ trend as clearly seen from $Q$ factor of Design 2 in Fig. 3. For Designs 1 and 3, the $N^5$ trend stops at smaller number of rings $N$ not shown in Fig. 3, and this manifests because such designs have larger dimensions than Design 2 hence the impact of losses is significant in Designs 1 and 3. Furthermore, Designs 2 and 3, accounting for losses for $N = 12$ rings, have almost the same $Q$ factor as seen in Fig. 3(a), even though Design 3 has a total length five times larger than that of Design 2 for the same $N$ (yet they have almost identical coupling parameters). In principle, this indicates that to attain a specific $Q$ factor using a lossy DBE CROW with the smallest possible area while keeping all the other CROW parameters fixed, it is preferable to utilize rings with the smallest dimensions.

We provide a comparative analysis for the three designs to show enhancement of the normalized Q factor defined as the ratio between the $Q$ factor of the CROW to the single ring resonator $Q$ factor. The latter is given by $Q_{0,\text{loss}} = \omega_{r,d} \tau_{ph} / (\alpha L_r - \ln(1-\kappa^2))$ [23] where $\omega_{r,d}$ is the angular frequency coinciding with the one of the CROW with DBE, $L_r = 2\pi R$ is the circumference of each ring resonator, $\tau_{ph}$ is the phase delay given by $\tau_{ph} = n_r L_r / c$ and $n_r$ is the effective refractive index of the ring resonator; while $\alpha$ is the waveguide power-attenuation constant. The respective single ring $Q$ factor of the three designs considered here for both lossless (denoted by $Q_0$ for the case when $\alpha = 0$) and lossy (denoted by $Q_{0,\text{loss}}$) are also reported in Table. 1. We show in Fig. 3(b) that the anomalous scaling of normalized $Q$ near the DBE depends on $Q_0$ as well as on the DBE parameters, i.e., is a function of $Q_0$ and the constant $\zeta$.

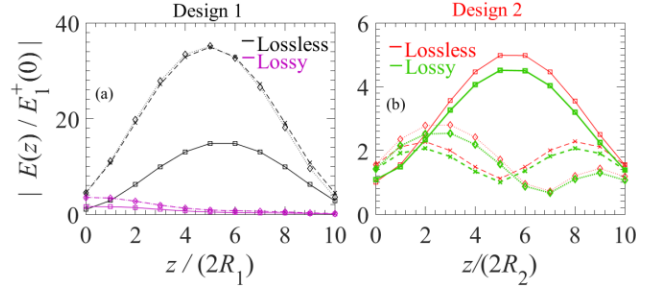

Fig.4. Electric field amplitudes at the unit cell boundaries of Design 1 and Design 2 in both lossless and lossy CROWs. Solid lines represent the normalized electric field in the waveguide ($|E_1(z)/E_1^+(0)|$), while dashed and dotted lines represent the normalized electric field inside the ring resonators ($|E_2(z)/E_1^+(0)|$ and $|E_3(z)/E_1^+(0)|$, respectively). Note that $z$ is normalized to the unit cell length of Design 2 which is $2R_2$.

Notice that in Design 1, the $Q$ factor of the single ring resonator (which could be considered simply as the $Q$ factor of a CROW with a single ring or $N=1$) is smaller than the other two designs (due to its larger value of $\kappa$). Yet, interestingly such configuration produces the highest possible $Q$ among the three designs for both lossless and lossy structures. For instance, for $N = 10$ rings, the $Q$ factor of Design 1 with losses is higher than that of Design 3 with losses and it is even higher than that of Design 2 without losses. This is attributed to having the smallest value of $\zeta$ (the flattest dispersion) as well as a large value of the stored energy in the DBE CROW for Design 1, see Fig. 4(a). We recall that the DBE resonance shows an unconventional standing wave profile mandated by giant field concentration in the center of the cavity [7,10]. By examining the field distribution in Designs 1 and 2 as shown in Fig. 4, we see that the resonance peak field exists as expected at the center region of the CROW ($z \sim L/2$) either inside the ring resonators (Design 1, Fig. 4a) or inside the straight waveguide (Design 2, Fig. 4b). In fact, for Design 1 the field is remarkably much higher than that of Design 2, that is why it has larger $Q$ factor despite having the smallest $Q_0$. In addition, the field is concentrated in the rings due to larger value of $\kappa$. Note that the field profile associated to Design 2 in the presence of losses maintains the ideal DBE resonance field profile; whereas in Design 1 the DBE resonance shape is largely perturbed. This mechanism also can be used to engineer the mode profile inside such CROW to control the impact of losses and to design highly sensitive sensors.

To further elucidate the anomalous scaling of CROWs with DBE, we compare their Q factors to that of other resonator designs without DBE: We compare with a conventional CROW made of coupled ring resonators *without* coupling to the straight waveguide; also we compare it with an optically large single ring resonator whose diameter $D_s$ equals the total length of CROW (i.e., $D_S = 2NR$), and finally we compare it with a design of a chain of cascaded *uncoupled* ring resonators (i.e. similar to the proposed CROW with waveguide but with vanishing coupling between the adjacent rings, i.e., with $\kappa' = 0$ ). In the *lossless* case, the conventional CROW $Q$ factor scaling is proportional to $N^3$ when operating at an RBE [2]. On the other hand, the large single ring resonator of radius $NR$ has a $Q$ factor equals to $NQ_0$ (i.e., exhibiting a linear growth of $Q$ with length). Hence, when the CROW's normalized $Q$ factor, i.e., $Q/Q_0$ (shown in Fig. 3(b)) exceeds *N*, the CROW's $Q$ factor surpasses that of the respective large single ring resonator. As such, $Q$ factor of Design 1, in the range shown in Fig. 3(b), is always higher than that of the respective large single ring resonator. Whereas the other two designs start to have higher $Q$ factor than their respective large single ring after 11 rings. The behaviour of the chain of cascaded uncoupled ring resonators, is identical to that of the large single ring resonator since $Q$ grows linearly with *N*.

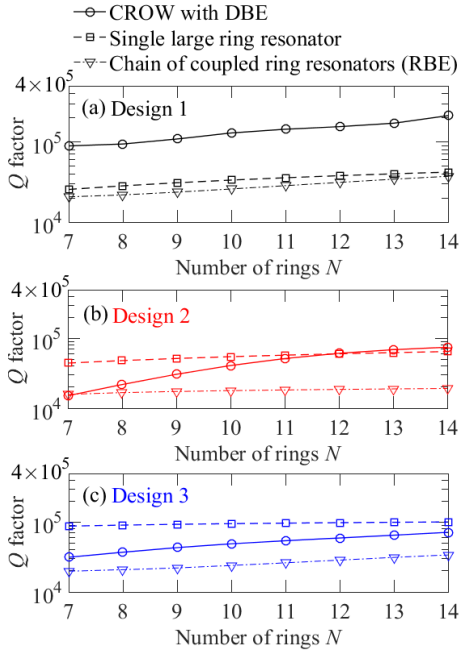

Fig. 5. *Q*-factor of lossy CROWs, for the three designs in Table. 1, represented by solid lines. Results are compared to the $Q$ of lossy single ring resonators having diameters equal to the respective CROW length (i.e. $D_s = 2NR_t$) represented by dashed lines. The dashed dotted lines with triangular markers represent $Q$ factor of the conventional CROW (chain of coupled ring resonators).

In the *lossy* case, the $Q$ factor of the CROW with DBE is compared to that of the other two resonators in Fig. 5; namely to the lossy chain of coupled ring resonators (conventional CROW), and to the respective lossy large single ring resonator. From Fig. 5 we observe that the $Q$ factor of the conventional CROW which operates at an RBE is always worse than the other designs. On the other hand, Design 1 with losses shows always better performance than all other designs, in the range shown in Fig. 5(a). Lossy Design 2 in Fig. 5(b) starts to show better $Q$ factor than the large single ring resonator when $N > 12$ rings, whereas the $Q$ of lossy Design 3 does not exceed the $Q$ of the single ring resonator till $N = 14$ rings as seen in Fig. 5(c).

We finally study the impact of perturbations on the DBE $Q$ factor of the CROW shown in Fig. 1. Indeed, during a microfabrication process structural, perturbations from the ideal design occur. Especially, the coupling parameters are dictated by tolerances in the gaps between adjacent rings and between rings and the straight waveguide. Let us assume that the values of $\kappa$ and $\kappa'$ in each unit cell of the N-rings CROW are varied within 5% change of their DBE design value in Table. 1 using a standard uniform probability density function. In other words, we assume a uniform distribution in the interval $[0.95\kappa, 1.05\kappa]$. which are within the limits of modern fabrication tolerances [24]. We perform sufficient random simulations within this interval and calculate the statistics: namely the average Q factor and standard deviation of the Q factor, namely $\sigma_Q$ as shown in Fig. 6. For the lossless case, we show that the standard deviation increases as the number of cells increases ($\sigma_Q$, is represented by vertical bars). Despite perturbations, the effect of growing $Q$ is still remarkable. For the lossy CROW designs, $\sigma_Q$ is very small, almost unnoticeable when compared to average values of *Q* (as seen from the zoomed inset in Fig. 6). In summary, the trend of *Q* versus *N* is almost independent of possible fabrication tolerances; thus, DBE is robust against some standard fabrication tolerances.

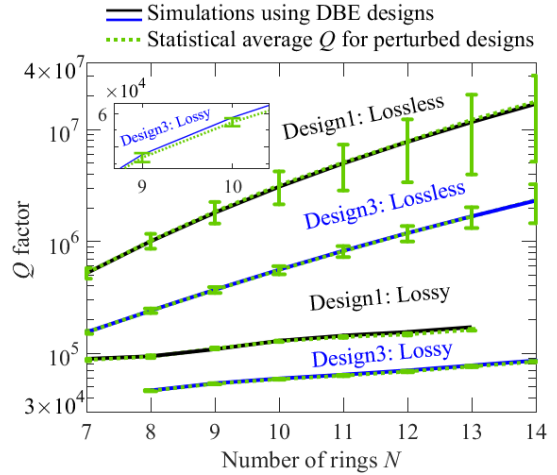

Fig. 6. Statistics of the effect of random perturbations on the $Q$ factor of Design 1 and Design 3 for both lossless and lossy cases when the structure is perturbed within 5% deviation in both $\kappa$ and $\kappa'$ form their DBE design values in Table 1. Dotted lines represent the statistical average of $Q$ factor while error bars denote the standard deviation $\sigma_Q$.

## CONCLUSION

We have demonstrated that the degenerate band edge (DBE), a fourth order EPD, occurs in a properly engineered CROW coupled to a waveguide. We have shown an unprecedented scaling of the *Q* factor with length even in the presence of losses. We have illustrated a very effective approach to enhance the *Q* at optical frequencies by properly engineer the coupling coefficients and dimensions of the proposed CROW. Importantly, the desired large *Q* values associated to the DBE resonance are shown to be robust against possible fabrication tolerances and could be readily detected in experiments. It is important to notice that certain designs of CROW with DBE show much larger *Q* factor than others, and we have explained how this depends on the parameter ζ, that is thus important to obtain the benefits of the CROW with DBE. The dependence of DBE resonance on structural/environment parameters could be further investigated to conceive novel extremely sensitive sensors.

## ACKNOWLEDGEMENT

This material is based upon work supported by the Air Force Office of Scientific Research under award number FA9550-15-1-0280 and under the Multidisciplinary University Research Initiative award number FA9550-12-1-0489 administered through the University of New Mexico.